\documentstyle[aps,preprint]{revtex} 
\draft

\begin{document}
\input psfig

\date{\today} \title{Possibility of p-wave pairing of composite 
fermions at $\nu=\frac{1}{2}$}
\author{K. Park,$^{a}$ V. Melik-Alaverdian,$^{b}$ N.E. Bonesteel,$^{b}$
and J.K. Jain$^{a}$}  
\address{$^{a}$ Department of Physics and Astronomy, State University
of New York, Stony Brook, New York 11794-3800\\ 
$^b$ National High Magnetic Field Laboratory and Department of Physics, Florida State
University, Tallahassee, Florida 32310}
\maketitle
\begin{abstract}
We find that for the pure Coulomb repulsion the composite Fermi sea at
$\nu=1/2$ is on the verge of an instability to triplet pairing of
composite fermions.  It is argued that a transition into the paired
state, described by a Pfaffian wave function, may be induced if the
short-range part of the interaction is softened by increasing the
thickness of the two-dimensional electron system. 
\end{abstract}
\maketitle

It has been over 10 years since the surprising discovery of an
even-denominator fractional quantum Hall effect (FQHE) at Landau level
(LL) filling fraction $\nu=5/2$ \cite{willett}.  In this state the
lowest ($n=0$) LL is filled for both up spins and down spins and the
effective filling factor of the first-excited ($n=1$) LL is 1/2.  In
an attempt to explain how this state was able to escape the usual `odd
denominator rule' of the FQHE, Haldane and Rezayi
\cite{haldanerezayi} proposed a trial wave function which described an
incompressible singlet state for a half-filled LL and argued
that this state might be stable at $\nu=5/2$. Despite some initial
experimental support for the non-fully-polarized nature of the state
in tilted field experiments \cite{eisenstein}, questions persisted
from the very beginning about whether this was in fact the correct
description of the $\nu=5/2$ state \cite{macdonald}.  Exact
diagonalization calculations\cite{chakraborty,morf} also indicate that
the true {\em Coulomb} ground state at $\nu=5/2$ is not spin singlet
even in the limit of zero Zeeman coupling. Greiter {\em et al.} \cite
{greiter} raised an alternative possibility in which the 5/2 FQHE
state is {\em fully polarized} and described by the Pfaffian wave
function proposed by Moore and Read \cite {mooreread}. This scenario
was recently further explored by Morf \cite {morf}.

In contrast, there has been significant progress in our understanding
of the physics of the compressible state at $\nu=1/2$ in terms of the
composite fermion (CF) theory, where composite fermions are electrons
bound to an even number of vortices in the many-body wave function
\cite{jain}.  According to this theory, interacting electrons in the
lowest LL are described in terms of composite fermions at an effective
magnetic field.  In particular, the FQHE for electrons can be viewed
as an effective {\it integer} quantum Hall effect for composite
fermions \cite {jain} and the {\em compressible} state at a
half-filled LL as a `metal' of composite fermions with a sharp Fermi
surface \cite{halperinleeread}.  A growing number of experiments have
confirmed the existence of such a `CF sea' at $\nu=1/2$
\cite{experiments}. Recently, Haldane and Rezayi \cite {HR} have
suggested that the true ground state at $\nu=1/2$ may actually be a
``weakly coupled" paired CF state for the Coulomb interaction, going
smoothly into the ``strongly coupled" Pfaffian state as the
pseudopotential $V_3$ is increased relative to its Coulomb value; the
CF sea appears as the $T>T_c$ normal state in this scenario.

Motivated by these issues, we have carried out a systematic study of
five different trial wave functions. Specifically we have considered
wave functions which describe the compressible spin singlet and spin
polarized CF sea states \cite{jain}, the incompressible spin singlet
Haldane-Rezayi \cite{haldanerezayi} and Belkhir-Jain
\cite{belkhir} states, and finally the incompressible spin polarized
Pfaffian state \cite{mooreread}. Our principal finding is that at
$\nu=1/2$ the Pfaffian state has an energy that is surprisingly close
to that of the fully polarized CF sea, and in fact, there is numerical
evidence that a transition to the former may take place as a function
of increasing thickness of the electron wave function perpendicular to
the plane of the two-dimensional electron system.

We have performed our simulations using Haldane's spherical geometry
\cite{haldane} in which $N$ electrons are confined to the surface of a
sphere of radius $R$.  A monopole at the center of the sphere produces
a radial field corresponding to $2Q$ flux quanta ($\phi_0=h/e$)
piercing the surface of the sphere. The one-body eigenstates are the
monopole harmonics \cite{wu} $Y_{Q,l,m}(\theta_i,\phi_i)$ where $l$
and $m$ are the angular momentum quantum numbers.  It is also
convenient to define the spinor coordinates $u_i = \cos
[\theta_i/2]\exp [i\phi_i/2]$ and $v_i = \sin [\theta_i/2]
\exp[-i\phi_i/2]$ where $\theta_i$ and $\phi_i$ are the usual
spherical coordinates.  For spin singlet states it will be assumed
that a particle is spin up if $i \le N/2$ and spin down if $i > N/2$.
For a given $\nu$ the relationship between the number of flux quanta
and the number of particles is $2Q = \nu^{-1}(N-1)-S$ where the
order-unity shift $S$ depends on the state being considered.

The CF states are the lowest LL projections of wave functions of the
form (Jastrow Factor) $\times$ (Slater Determinant):
\begin{eqnarray}
\psi = {\cal P}_{LLL} \Phi_1^2 \Phi=\Phi_1^2 \tilde \Phi\;.
\label{cfwf}
\end{eqnarray}
Here, $\Phi_1 = \prod_{i < j} (u_i v_j - v_i u_j)$, ${\cal P}_{LLL}$
is the lowest LL projection operator, and $\Phi$ is the $N \times N$
Slater determinant state of electrons at effective flux $q=Q-N+1$,
made of one-body eigenstates $Y_{q,l_i,m_i}(\theta_j,\phi_j)$.  The
lowest LL projection is carried out following the procedure devised by
Jain and Kamilla \cite{jain-projection}, which amounts to replacing
the monopole harmonics in $\Phi$ by the {\em projected} monopole
harmonics $\tilde Y_{q,l_i,m_i}(\theta_j,\phi_j)$, defined by
\begin{eqnarray}
\tilde Y_{q,l,m} = {\cal J}_i^{-1} {\cal P}_{LLL}{\cal J}_i Y_{q,l,m}
\end{eqnarray}
where ${\cal J}_i = \prod_{k(\ne i)} (u_i v_k - v_i u_k)$. This
changes $\Phi$ into $\tilde \Phi$ to give the last equality of
Eq.~(\ref{cfwf}).  Explicit analytic expressions for $\tilde
Y_{q,l,m}$ are given in Ref. \cite{jain-projection}.

The spin polarized and spin singlet CF sea states have the form
\begin{eqnarray}
\psi = {\cal P}_{LLL} \Phi_1^2 \Phi_{F.S.}
\end{eqnarray}
where $\Phi_{F.S.}$ is the $N \times N$ Slater determinant ground
state of electrons at ``zero effective flux" ($q=0$), chosen
appropriately to be either fully polarized or unpolarized.  For the
spin-polarized case states of progressively higher $l$ values are
filled until a closed shell configuration is reached.  This occurs
when $N = p^2$ where $p$ is an integer and results have been obtained
for $N=4,9,16,25$ and 36.  For the spin-singlet case each state is
doubly occupied by a spin up and spin down composite fermion. The
closed shell configurations then occur when $N=2 p^2$ and results have
been obtained for $N=8,18,32$ and 50 electrons.

The Haldane-Rezayi state \cite{haldanerezayi} is given by $\Phi_1^2 \;
\det M$, where $M$ is the $N/2 \times N/2$ matrix with components
$M_{ij} = (u_i v_{j+N/2} - v_i u_{j+N/2})^{-2}$ where $i,j=1,\cdots
N/2$. The other incompressible spin-singlet state we have considered
is the Belkhir-Jain state \cite{belkhir}, $\Phi_1 \Phi_{1,1} \Phi_2$,
where $\Phi_{1,1}$ is the wave function of the lowest LL with both
spin up and spin down states fully occupied, and the matrix $\Phi_2$
is an $N \times N$ Slater determinant corresponding to two filled LLs,
at effective flux $2q = 2Q-(3N/2-2) = (N-4)/2$, constructed as if the
electrons were spinless.  The lowest LL projection is carried out by
writing it as $\Phi_{1,1} \Phi_1^{-1} {\cal P}_{LLL} \Phi_1^2 \Phi_2 =
\Phi_{1,1}\Phi_1 \tilde \Phi_2$.

Finally, the Pfaffian state is a spin polarized FQHE state
\cite{mooreread} which can be written
\begin{eqnarray}
\psi_{Pf} = \Phi_1^2\; {\rm Pf} M
\end{eqnarray}
where ${\rm Pf} M$ is the Pfaffian of the $N \times N$ antisymmetric
matrix $M$ with components $M_{ij} = (u_i v_j - v_i u_j)^{-1}$.  As
pointed out by Greiter {\it et al.} \cite{greiter}, ${\rm Pf} M$ is a
real space BCS wave function and so $\psi_{Pf}$ can be viewed as a
$p$-wave paired quantum Hall state.

All of these wave functions, with the exception of the Pfaffian state,
are of the form (Jastrow Factor) $\times$ (Determinant), and can be
studied by standard variational Monte Carlo methods. For the Pfaffian
state the identity $|{\rm Pf} M|^2 = |{\rm det} M|$ (up to an
irrelevant normalization factor) can be used, and, again, standard
variational Monte Carlo techniques can be applied.  For each of these
states the correlation energy per particle $E = \frac{n}{2}\int (g(r)
-1 ) V(r) d^2r$, where $n$ is the carrier density, $g(r)$ is the pair
correlation function and $V(r)$ is the electron-electron interaction,
has been calculated for systems containing up to 50 particles and the
results extrapolated to the $N \rightarrow \infty$ limit.

\underline{$\nu=1/2$}. --- 
The extrapolated Coulomb energies per particle obtained for the five
states we have studied are given in Table I.  Results are in units of
$e^2/\epsilon l_0$ where $\epsilon$ is the dielectric constant and
$l_0 = (\hbar c/eB)^{1/2}$ is the magnetic length.  At $\nu=1/2$, the
lowest energy state is the singlet CF sea state, closely followed by
the polarized CF sea and the Pfaffian states.  The Haldane-Rezayi and
Belkhir-Jain states have significantly higher energies and shall not
be considered further.

The difference between the energies (per particle) of the polarized
and unpolarized CF sea states is $\approx 0.004 e^2/\epsilon l_0$.  In
a model of noninteracting composite fermions with an effective mass
$m^*_p$ (the ``polarization mass"), this is equated to the kinetic
energy difference between the polarized and unpolarized CF seas to
give
\begin{equation}
\frac{1}{8}\frac{\hbar eB}{m_p^* c}=0.004 \frac{e^2}{\epsilon l_0}\;.
\end{equation}  
In contrast, the ``activation mass" $m^*_a$ of composite fermions,
defined by equating the excitation gap to an effective cyclotron
energy gives $\frac{\hbar eB}{m_a^* c}=0.32 \frac{e^2}{\epsilon l_0}$,
implying that $m_p^*/m_a^*\approx 10$, roughly consistent with the
result in Ref. \cite {park}. For typical magnetic fields, the actual
ground state will be a partially polarized CF sea.

\begin{figure}
\centerline{
\psfig{figure=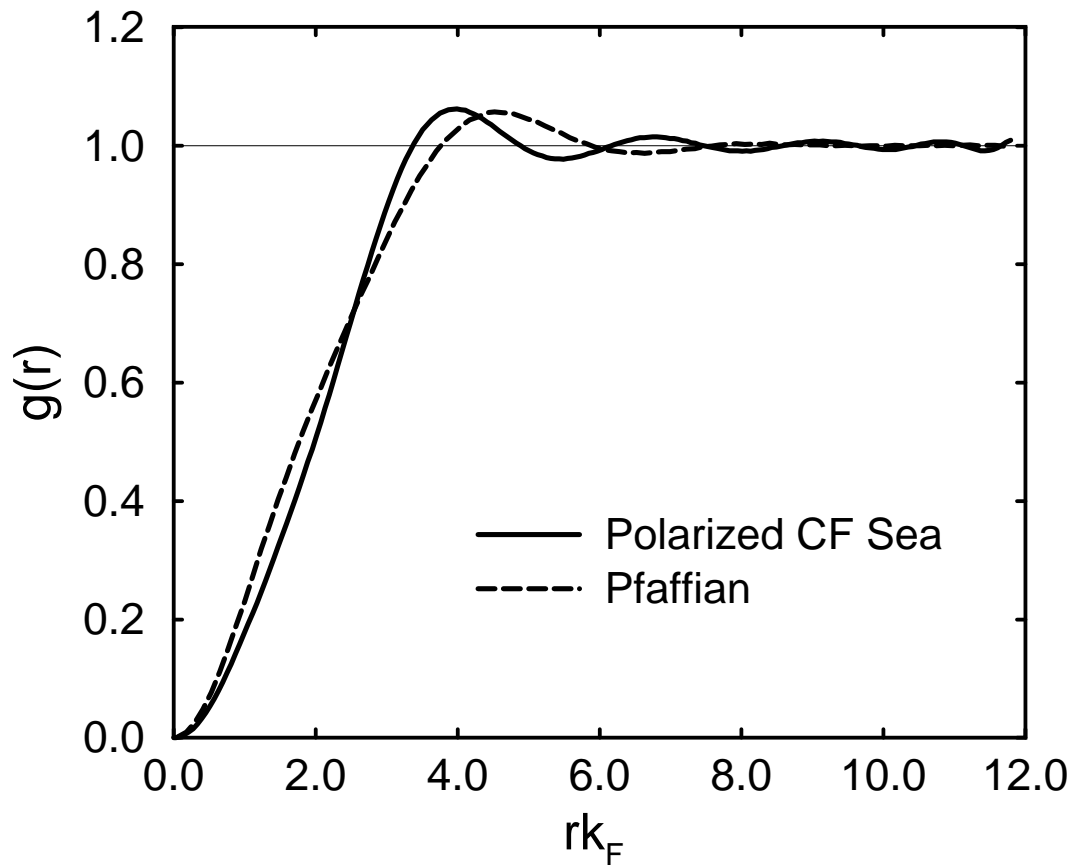,height=5in,angle=-90}}
\ 
\
\caption{Pair correlation functions for the Pfaffian and spin polarized
composite fermi sea wave functions.  Results are for systems with 36
electrons.}
\label{pcf}
\end{figure}

It is remarkable that the Pfaffian and polarized CF sea states are so
close in energy given their qualitatively different natures. This
difference can be seen in Fig.~(\ref{pcf}) in which the pair
correlation functions for these two states are shown for a system with
36 electrons plotted as a function of $r k_F$ where $r$ is the chord
distance on the sphere and $k_F = l_0^{-1}$ is the Fermi wave vector
of the polarized CF sea. For the pair correlation function of the CF
sea one sees $2k_F$ oscillations which fall of as a power law for
large $r$ \cite{kamilla,rezayiread} consistent with the existence of a
sharp `Fermi surface' of composite fermions. Similar oscillations are
strongly damped for the pair correlation function of the Pfaffian
state which presumably approaches the asymptotic value of unity
exponentially with increasing $r$.  In going from the polarized CF sea
to the Pfaffian state there is an increase in $g(r)$ for small $r$
which we interpret as a signature of real space pairing correlations.

\begin{figure}
\centerline{ \psfig{figure=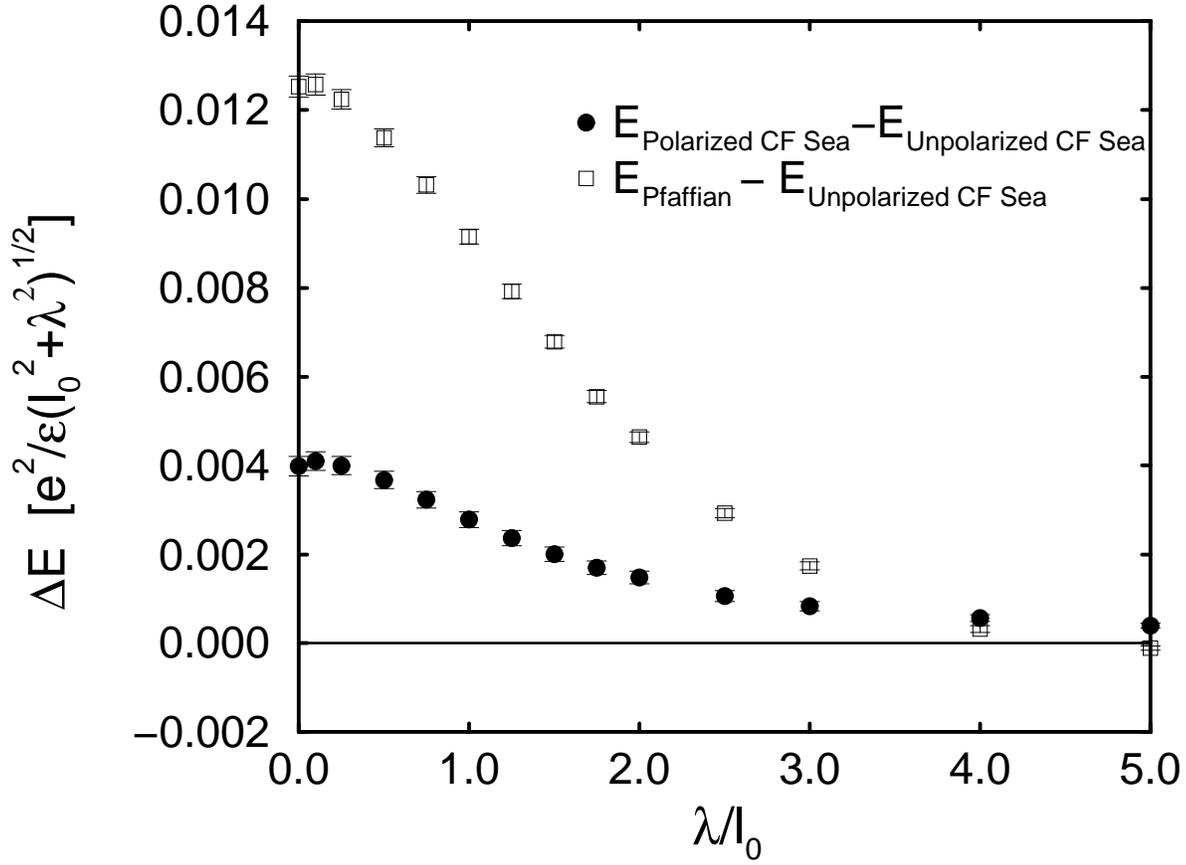,height=5in,angle=-90}}
\
\
\caption{Energy difference between the Pfaffian state and the 
unpolarized composite Fermi sea (open squares) and between the
polarized and unpolarized composite Fermi sea (solid circles) vs.
$\lambda/l_0$ where $\lambda$ characterizes the thickness of the
two-dimensional electron system.  Energy differences are given in
units of $e^2/\epsilon(l_0^2 + \lambda^2)^{1/2}$.}
\label{thick}
\end{figure}

The fact that the correlation energy of the Pfaffian state is close to
that of the CF sea, which in turn is believed to be an excellent
representation of the true Coulomb ground state in the lowest LL,
makes it plausible that the former may be relevant for an interaction
not too different from the pure Coulomb interaction.  [Strictly
speaking, a variational study is too crude to distinguish between
states with small energy differences and cannot rule out the
possibility that even for the Coulomb interaction the true ground
state is paired, as argued in \cite {HR}, but we will assume this not
to be the case in view of the facts that no FQHE is observed at
$\nu=1/2$ and that there is experimental evidence for a Fermi sea at
$\nu=1/2$.]  It would be of interest to explore if a transition from
the compressible CF sea to the Pfaffian may be induced at $\nu=1/2$ by
tuning some experimentally controllable parameter, e.g.  the thickness
of the two-dimensional electron system, which alters the detailed form
of the interaction potential.  To investigate this possibility we have
modeled the effect of finite thickness by replacing the pure Coulomb
interaction by the effective interaction
\cite{zhang},
\begin{eqnarray}
V(r) = \frac{e^2}{\epsilon \sqrt{r^2 + \lambda^2}}.
\end{eqnarray}
The energy differences between the unpolarized CF sea and the
polarized CF sea and the Pfaffian state are plotted as a function of
$\lambda/l_0$ in Fig.~(\ref{thick}).  To account for the reduction of
the characteristic energy scale with increasing thickness the energy
difference is given in units of $e^2/\epsilon(l_0^2 +
\lambda^2)^{1/2}$. For $\lambda \agt 4 l_0$ we find that the Pfaffian
has lower energy than the fully polarized CF sea, and for $\lambda
\agt 5 l_0$ its energy is below even that of the unpolarized CF sea.
Thus we expect that, independent of whether the CF sea is fully or
partially spin polarized, it should be possible to induce a transition
to the Pfaffian state by increasing the thickness.

\begin{figure}
\centerline{
\psfig{figure=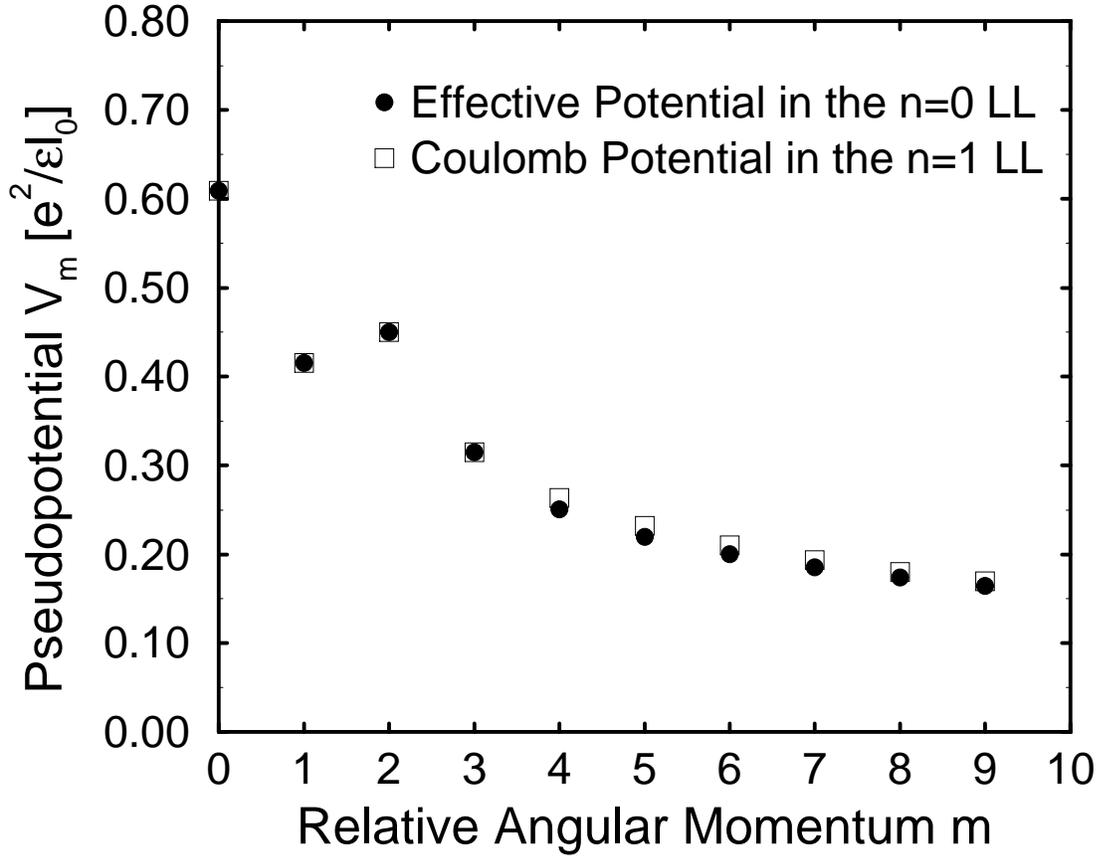,height=5in,angle=-90}}
\
\
\caption{
Haldane pseudopotentials for the effective potential $V_{eff}$
discussed in the text in the $n=0$ Landau level (solid circles) and
for the Coulomb potential in the $n=1$ Landau level (open squares).}
\label{pseudo}
\end{figure}

\underline{$\nu=5/2$}. ---
While it is conceptually straightforward to promote the above wave
functions to the first excited LL, a computation of the energy is
difficult due to a lack of an explicit form. We instead proceed by
working with an {\em effective} interaction in the lowest LL, which is
equivalent to the Coulomb interaction in the second LL.  This
interaction is derived by requiring that its pseudopotentials in the
lowest LL are the same as the pseudopotentials of the Coulomb
interaction in the second LL. We remind the reader that the Haldane
pseudopotentials $V_m$ \cite{haldane} are simply the correlation
energies of pairs of particles in a given LL with relative angular
momentum $m$.  We have used the following effective potential,
\begin{eqnarray}
V_{eff}(r) = \frac{e^2}{\epsilon}
\left(
\frac{1}{r} + a_1 e^{-\alpha_1 r^2} + a_2 r^2
e^{-\alpha_2 r^2}
\right)\;,
\end{eqnarray}
The parameters $a_1$, $a_2$, $\alpha_1$, and $\alpha_2$ have been
fixed by requiring that the first four pseudopotentials of
$V_{eff}(r)$ for $n=0$ be exactly equal to the first four
pseudopotentials of the Coulomb repulsion for $n=1$ (the results of
this procedure are $a_1 = 117.429$, $a_2 = -755.468$, $\alpha_1
=1.3177$ and $\alpha_2 = 2.9026$).  The remaining pseudopotentials are
asymptotically correct because $V_{eff}(r) \simeq e^2/\epsilon r$ for
large $r$.  The pseudopotentials for the Coulomb interaction in the
$n=1$ LL and for $V_{eff}$ in the $n=0$ LL are shown in
Fig.~(\ref{pseudo}).  It can be seen clearly that the effective
potential $V_{eff}$ does an excellent job of characterizing the
Coulomb interaction in the $n=1$ LL.

The energies of various wave functions at $\nu=5/2$ are now
straightforwardly computed as before, with the results also shown in
Table I. The lowest energy state here is the Pfaffian, with the
spin-polarized CF sea having only slightly higher energy and all three
singlet wave functions having much higher energy.  The correlation
energy per particle we obtain for the Pfaffian, $-0.362(2)
e^2/\epsilon l_0$, is remarkably close to the extrapolated exact
diagonalization calculations of Morf\cite {morf} of $-0.366
e^2/\epsilon l_0$ [although it should be noted while comparing these
numbers that the $V_{eff}(r)$ used in our calculations is slightly
less repulsive than the actual Coulomb interaction (Fig.~3)].  We
stress, however, that while the above variational calculations make
the Pfaffian state plausible, more work will be required to
definitively establish its relevance to the true FQHE state at
$\nu=5/2$.  This is in contrast to the situation in the lowest LL FQHE
where the CF wave functions have been found to have close to 100\%
overlap with the exact ground states.

The state which lies lowest in energy for a given potential is
determined by the relative strengths of the various pseudopotentials.
The superiority of the Pfaffian wave function over the fully polarized
CF sea for large thickness at $\nu=1/2$ and for zero thickness at
$\nu=5/2$ are somewhat analogous: in both cases, the short-range part
of the interaction is suppressed relative to pure Coulomb interaction.
Also, the tendency for full spin polarization in the second LL may be
attributed to the relatively high Coulomb energy cost of having pairs
of particles with relative angular momentum $m=2$ (Fig.~3).

To summarize, we have presented a variational Monte Carlo study of
several wave functions both for $\nu=1/2$ and $\nu=5/2$.  Of the
states we have considered we find that for the pure Coulomb repulsion,
the spin (un)polarized CF sea is the ground state at (zero) large
Zeeman coupling at $\nu=1/2$, and the incompressible spin-polarized
Pfaffian state lies lowest at $\nu=5/2$.  The possibility of a
transition at $\nu=1/2$ from the CF sea to the Pfaffian state as a
function of the thickness of the two-dimensional electron system has
been investigated.

The authors would like to thank F.D.M. Haldane, E. Rezayi, and Z. Ha
for useful discussions.  This work was supported in part by the
National Science Foundation under grant no. DMR-9615005 and by the US
Department of Energy under grant no.  DE-FG02-97ER45639.  NEB
acknowledges support from the Alfred P.  Sloan Foundation.  A grant of
computing time on the SGI Power Challenge cluster at the NCSA,
University of Illinois, Urbana-Champaign is acknowledged.

\begin{table}
\caption{
Correlation energies of the five states considered in this paper for
both $\nu=1/2$ and $\nu=5/2$.  All results have been extrapolated to
the thermodynamic limit and are given in units of $e^2/\epsilon l_0$}
\begin{tabular}{cccccc}
$\nu$  & {\rm Pfaffian} &
{\rm Composite Fermi Sea} &
{\rm Composite Fermi Sea} &
{\rm Haldane-Rezayi} &
{\rm Belkhir-Jain} \\
 & (Polarized) &
 (Polarized) &
 (Singlet) &
 (Singlet) &
 (Singlet) \\
\tableline
$\frac{1}{2}$ & -0.45694(17)  & -0.46557(6)  & -0.46953(7)  &
-0.31470(33) & -0.41691(29) \\
$\frac{5}{2}$ & -0.3621(22)   &  -0.34919(54)   & -0.29517(30)  &
-0.3032(31) 
& -0.2872(16) \\
\end{tabular}
\end{table}

\end{document}